\documentstyle[12pt,epsf]{article}
\date{\today}
\newcommand{\vkappa}{{\mbox {\boldmath$\kappa$}}}
\newcommand{\vq}{{\bf q}}
\newcommand{\vrho}{{\mbox {\boldmath$\varrho$}}}

\begin{document}

\title{Physical mechanism of the linear beam--size effect at colliders}

\author{
K.~Melnikov\\{\em Institut f\"ur
Physik,Universit\"{a}t Mainz}\thanks{ D 55099 Germany, Mainz,
Johannes Gutenberg Universit\"{a}t, Institut f\"{u}r Physik,
THEP, Staudinger weg 7; e-mail: melnikov@dipmza.physik.uni-mainz.de}
\\and\\
G.~L.~Kotkin~and~V.G.~Serbo\\{\em
Novosibirsk State University}\thanks{
Novosibirsk State University, 630090, Novosibirsk, Russia;
e-mail: kotkin@phys.nsu.nsk.su; serbo@math.nsk.su}}

\maketitle

\begin{abstract}
We present qualitative but precise description of the linear beam--size
effect predicted for the processes in which unstable but long--living
particles collide with each other. We derive  physically pronounced
equation for the events rate which proves that the linear beam--size
effect corresponds to the scattering of one beam of particles on the
decay products of the other. We compare this linear beam--size effect with
the known logarithmic beam--size effect measured in the experiments
on a single bremsstrahlung at VEPP-4 and HERA.
\end{abstract}
\vspace{0.5cm}
\begin{center}
MZ-TH/96-11
\end{center}

\newpage

\section{Introduction}

Let us consider a process with unstable but long--living
particle(s) in the initial state. Calculating corresponding cross
section, one can meet a singularity in the physical region
related to the appearance of real (not virtual) particles in an
intermediate state.  The appearance of such singularities leads
to a divergent cross section.  To perform calculations in this
case a careful analyses of the physical set--up is needed.

Such situation takes place for a number of processes at muon colliders,
for instance,
$\mu^+\mu^- \to e\bar \nu_e X$, $\mu^+\mu^- \to e\nu_\mu X$, ...
which have been recently discussed in refs. \cite{Ginz, MS}.
To be concrete, below we consider the process
\begin{equation}
\mu^-\mu^+ \to e\bar \nu_e W^+\,.
\label{1}
\end{equation}
One of the Feynman diagrams of this process which contains
the abovementioned singularity (the so--called $t$-channel
singularity in the physical region) is presented in the
Fig.1. This diagram can be viewed as a sequence of two processes:
\begin{equation}
\mu^- \to e\bar \nu_e \nu_\mu
\label{2}
\end{equation}
and
\begin{equation}
\nu _\mu \mu ^+ \to W^+
\label {2.5}
\end{equation}
both of which can occur for the {\it real} muonic neutrino.
Therefore, there exists a region of the final phase space
corresponding to the reaction (\ref {1}) where denominator of the
propagator of the muonic neutrino $q^2$ can be equal to zero
\footnote{For the appearance of such singularity it is necessary
that both of the processes (\ref {2}, \ref {2.5}) have a possibility to
occur.}.

\begin{figure}[htb]
\epsfxsize=8cm
\centerline{\epsffile{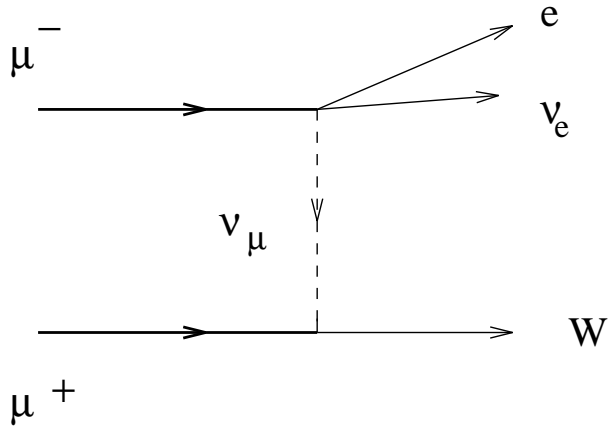}}
\caption[]{
The Feynman diagram for the reaction $\mu ^- \mu ^+ \to e \bar
\nu _e W^+ $ which has the $t$--channel singularity
and gives the leading contribution in the region
of small $|q^2|$.
}
\end{figure}

In the paper \cite{MS} it has been shown that accounting for the
finite sizes of the colliding beams gives a finite cross section for the
processes with the $t$-channel singularity in the physical region.
This result has been obtained on the base of the
results of the
paper \cite{Klass} where a general technique for taking into account the
finite sizes of the colliding beams
has been worked out (for the review
see \cite{Obzor}). The effective cross section obtained in
\cite{MS}
\begin{equation}
d\sigma = {dN\over L}
\label{3}
\end{equation}
(where $N$ is the number of events and $L$ is the
luminosity of the $\mu^+\mu^-$ collisions) can be written in the
form
\begin{equation}
d\sigma = d\sigma_{\mbox{\scriptsize{st}}}+
d\sigma_{\mbox{\scriptsize{nst}}}.
\label{4}
\end{equation}
The cross section $d\sigma_{\mbox{\scriptsize{st}}}$
by definition corresponds  to the wide region of the phase space
defined by the inequality
$q^2 < -m^2$ where
$m$ is the muon mass. This part of the cross section
can be calculated by means of the standard rules of the
relativistic scattering theory.
For the discussed process the result of this calculation
can be found in \cite {MS}.
The cross section
$d\sigma_{\mbox{\scriptsize{nst}}}$ corresponds to the
narrow region of the final phase space where
\begin{equation}
-m^2 < q^2 < t_0 , \;\;\; t_0>0
\label{5}
\end{equation}
which includes the point $q^2=0$. It turns out \cite {MS} that the main
contribution to this cross section is given by the pole of the
$\nu_\mu$ propagator and that this contribution is proportional
to the transverse size of the colliding beams $a$:
\begin{equation}
d\sigma_{\mbox{\scriptsize{nst}}} \, \propto \, a.
\label{6}
\end{equation}
It is {\it the linear beam--size effect}.

In the present paper we show that the non--standard piece of the
cross section $d\sigma_{\mbox{\scriptsize{nst}}}$ is defined by
the sequence of processes (\ref {2}), (\ref {2.5}) for real
muonic neutrinos. Such  interpretation of the results of ref.
\cite{MS} leads, in particular, to a more clear understanding of
its  region of applicability.

In view of the realistic situation  at future muon colliders,
we assume that longitudinal $l$ and transverse $a$
sizes of the colliding beams satisfy the
inequalities:
\begin{equation}
l \ll \gamma  \,c\,\tau, \;\;\;\;\;\; \gamma = {E\over mc^2},
\label{8}
\end{equation}
\begin{equation}
a \ll c\, \tau
\label{9}
\end{equation}
where $\tau$ is the muon life time in its rest frame, $c\, \tau =
660$ m. Due to the condition (\ref{8}) the number of muons
can be thought to be a constant during the time which is required
for the
colliding beams to cross each other. Note,
that conditions (\ref{8})--(\ref{9}) are perfectly fulfilled
for the  projects of muon colliders  which are discussed
in the literature \cite{Palm}. Using the beam parameters from
ref. \cite {Palm} and the total energy $2E=100$ GeV one gets:
\begin{equation}
{a\over c\tau} \sim {l\over \gamma c \tau} \sim 10^{-8}\,.
\label{10}
\end{equation}

Subsequent part of the paper is organized  as follows: in the next
section we give a simple qualitative description of the linear
beam--size effect; section 3 is devoted to the direct calculation
of the number of events corresponding to the $\nu_\mu \mu^+$
collisions; in section 4 we transform the result derived in
ref. \cite {MS} for $\sigma_{{\rm nst}}$ to the rest frame
of the $\mu ^-$ beam and compare it with the results of section
3; in section 5 we discuss a case when
conditions (\ref{8})--(\ref{9}) are not satisfied; our
conclusions are presented in  section 6.

\section{Qualitative description of the linear beam--size effect}

We consider a collision of a single muon $\mu^-$ with the
beam of muons of the positive charge $\mu^+$.
In this analysis it is convenient to use the rest
frame of the $\mu^-$. Being unstable, this muon
is surrounded by a ``cloud'' of
$\nu_\mu$'s which appear in the $\mu^-$ decay
(certainly, there are also  $\bar \nu_e$ and $e$ ``clouds'',
but they are not interesting for us at the moment).
The density of neutrinos in this cloud decreases as $1/r^2$ with the
growth of the distance $r$ from the $\mu^-$, its time dependence
corresponds to the exponential decay law and the angular
distribution is isotropic. Therefore, one finds
\begin{equation}
n_\nu ({\bf r},t) = {\theta (t-r/c) \over 4\pi c \tau r^2} \exp
\left(-{t-r/c \over \tau}\right).
\label{11}
\end{equation}
This density is normalized by the condition
$$
\int \limits _{0}^{\infty} \; c\, n_\nu ({\bf r}, t)\, 4\pi r^2\,dt = 1
$$
which means that the total number of $\nu_\mu$'s which cross a
sphere of radius $r$ is equal to unity.

We shall see below that the main contribution to the $\sigma
_{\mbox{\scriptsize{nst}}}$ comes from the distance $r \sim a$.
The typical time of collision is therefore $\Delta t \sim a/c$.
If the collision occurs at the time $t$ which satisfies the
inequality $\Delta t  \ll t \ll \tau$, the neutrino density can
be taken to be time independent in the collision region (these
assumptions are justified by the obtained result):
\begin{equation}
n_\nu ({\bf r}, t)\, = {1\over 4\pi c \tau r^2}.
\label{12}
\end{equation}

The distribution of the neutrinos over
impact parameters $\varrho$ of the $\mu^+\mu^-$ collisions
\footnote{
The $z$-axis is antiparallel to the momentum ${\bf p}_2$ of the
$\mu^+$\, .} is given by
\begin{equation}
dN_\nu = {d^2 \varrho \over 4\pi c \tau}\;
\int \limits _{-\infty}^{+\infty }\; {dz\over z^2+ \varrho^2}= {1\over
4c\tau} {d^2 \varrho \over \varrho}, \;\;\;\;\; {\bf r} =(\vrho,
z)\,.
\label{13}
\end{equation}
We note that the main contribution to this integral is
given by the longitudinal distances of the order of
$$
|z| \stackrel{<}{\sim} \varrho.
$$
The number of muonic neutrinos which collide with the
$\mu ^+$ beam of the radius $a$ is therefore:
\begin{equation}
N_\nu \sim \int \limits_0^a\; {d^2 \varrho \over 4c\tau \varrho}
= {\pi \over 2} {a\over c\tau}.
\label{14}
\end{equation}
The non--standard piece of the cross section of the process
(\ref {1}) is proportional to the number of neutrinos (\ref {14})
and can be estimated as:
\begin{equation}
\sigma _{\mbox{\scriptsize{nst}}} \sim {a\over c\tau}\; \langle
\sigma_{\nu \mu \to W} \rangle
\label{15}
\end{equation}
where $\langle \sigma_{\nu \mu \to W} \rangle$ corresponds to the
cross section of the process (\ref {2.5}) averaged over
effective $\nu_\mu$ spectrum \footnote {The exact form of this spectrum can
be found in ref. \cite {MS}.}.

The estimate (\ref{15}) corresponds to the
$\sigma_{{\rm nst}}$ obtained in \cite{MS}. The exact calculation is
presented in the next section.

\section{Direct calculation of the number of events in
$\nu_\mu \mu^+$ collisions}

Having performed the estimate, we proceed further with the exact
calculation. We perform it in the rest frame of the $\mu^-$ beam
in which neutrino density $n_\nu$ has the simple form
(\ref{12}). For simplicity, we neglect  energy and angular spread
of the particles in the muon beams.

The number of events for the $\nu_\mu \mu^+ \to W^+$ process can
be written in the form
\begin{equation}
dN_{\nu \mu \to W} = dL_{\nu \mu}(\omega)\, \sigma_{\nu \mu \to
W} (s_{\nu \mu}).
\label{16}
\end{equation}
Here $dL_{\nu \mu}(\omega)$ is the spectral luminosity of
$\nu_\mu \mu^+$ collisions. The cross section $\sigma_{\nu \mu
\to W}$ depends on the quantity \footnote{Below we use a system
of units where $\hbar =1$ and $c=1$.}
$$
s_{\nu \mu}= (q+p_2)^2 = 2\omega E_2 \, (1+\cos{\theta}) \,+\,
m^2
$$
where $\omega$ and $\theta,\,\varphi$ are the energy and the
escape angles of $\nu_\mu$, and $E_2$ is the energy of $\mu^+$.

For unpolarized muons the energy distribution of the $\nu_\mu$ does
not depend on the neutrino escape angles and has the well--known
form \cite {Okun}
\begin{equation}
w(\omega) d\omega = 16 \left({\omega \over m}\right)^2\, \left(3-
 {4\omega\over m}\right) \, {d\omega \over m}.
\label{17}
\end{equation}

Let $n_1({\bf r}_1)$ and $n_2 ({\bf r}_2, t)$ be the particle
densities of $\mu^-$ and $\mu^+$ beams respectively. In
accordance with eq. (\ref{12}) the density of the neutrinos in the point
${\bf r}_2$ is defined by the muons $\mu ^-$ decaying at the point
${\bf r}_1$ by the following equation:
\begin{equation}
dn_\nu ({\bf r}_2) = {n_1({\bf r}_1) \, d^3r_1 \over 4\pi c \tau
({\bf r}_2 - {\bf r}_1)^2}.
\label{18}
\end{equation}
Therefore, the spectral luminosity of $\nu_\mu \mu^+$ collisions
is equal to (see, for example, \cite{LL} \S 12)
\begin{equation}
dL_{\nu\mu}(\omega)= w(\omega) d\omega \; v \; dn_\nu ({\bf r}_2)
n_2 ({\bf r}_2, t)\, d^3r_2 dt
\label{19}
\end{equation}
where
\begin{equation}
v= {qp_2 \over \omega E_2} = 1+ \cos{\theta}.
\label{20}
\end{equation}

Note that the vector ${\bf r}_2 -{\bf r}_1$ is parallel
to the neutrino momentum ${\bf q}$:
\begin{equation}
({\bf r}_2 -{\bf r}_1) \; \parallel \; {\bf q}.
\label{21}
\end{equation}

In the next section we shall show that the result for $dN_{{\rm
nst}}$ obtained in \cite{MS} completely
coincides with eqs.
(\ref{16})--(\ref{20}).

\section{The number of events $dN_{{\rm nst}}$ in the rest frame
of the $\mu^-$ beam}

In this section we follow the line of reasoning of ref. \cite
{MS}. The only difference is that we switch to the rest frame of
the $\mu ^-$ beam. It turns out that this reference frame  is
quite helpful for the physical interpretation of the results
obtained in ref. \cite {MS}.

We use the following notations (see Fig. 1):
$s=(p_1+p_2)^2=4E^2$ is the square of the total energy in the
center of mass frame, $\Gamma = 1/\tau$  is the  muon
width, $p_1^2 =p_2^2 =m^2$, $p_3+p_4$ is the
4-momentum of the final $e^- \bar \nu _e$ system,
$y=(p_3+p_4)^2/m^2 $, $q = p_1-p_3 -p_4=(\omega, {\bf q})$ is the
momentum transfer in the $t$-channel and $x=qp_2 /p_1p_2$. The
square of the momentum transfer in the $t$-channel is equal to
\begin{equation}
q^2 = -{{\vq} _\bot {}^2\over 1-x} + t_0
\label{22}
\end{equation}
where  ${\vq} _\bot$ is the component of the momentum ${\vq}$
which is transverse to the momenta of the initial muons and $t_0$
is the maximal value of $q^2$
\begin{equation}
\max {\{ q^2 \} } \, =\,t_0 \, =  \, { x(1-x-y)\over 1-x} \, m^2.
\label{23}
\end{equation}
Note, that
\begin{equation}
t_0\, >\, 0 \;\;\; \mbox{for} \;\;\; y < 1-x.
\label{24}
\end{equation}

As is explained in the Introduction, usual approximations in
the scattering theory (based on the collisions of the plane
waves) result in the divergent cross section in this case. In contrast to
this approximation, we use the formalism of the colliding wave
packets (beams of particles) of macroscopic, but finite sizes
which has been developed in ref. \cite{Klass}. In the
overwhelming majority of experiments both descriptions provide
the same result. However, in a number of cases the usual
approximation is insufficient and consideration of the
colliding wave packets is mandatory --- see description of the
experiments on a single bremsstrahlung at VEPP-4 \cite {IYAF} and
at HERA \cite {HERA} and the review \cite{Obzor}. The same can be
stated about discussed processes with the singularities caused by
real particles in the intermediate states ($t$--channel
singularities in the physical region). The breakdown of the
standard calculations clearly indicates inapplicability of
the standard approximations in the scattering theory for these
cases.

Let us remind that in the standard approach the number of events
$N$ is the product of the cross section $\sigma $ and the
luminosity $L$:
\begin{equation}
dN=d\sigma~L,~~~ d\sigma \propto |M|^2, ~~~
L=v_{12}\int n_1({\bf r},t)~n_2({\bf r},t)d^3rdt
\label{25}
\end{equation}
where $v_{12}=|{\bf v}_1-{\bf v}_2| = 2$ for the head--on
collision of the ultra--relativistic beams. The quantities
$n_i({\bf r},t)$ are the particle densities of the beams.

The transformation from the plane waves to the wave packets
results in the following changes. The squared matrix element
$|M|^2$ with the initial state in the form of the plane waves with
the momenta ${\bf p_1}$ and ${\bf p_2}$ transforms to the product
of the matrix elements $M_{fi}$ and $M^*_{fi'}$ with
different initial states:
\begin{equation}
d\sigma \propto |M|^2 ~~ \to ~~ d\sigma ({\mbox {\boldmath
$\kappa $}}) ~\propto ~ M_{fi}M_{fi'}^{*} \, .
\label{26}
\end{equation}
Here the initial state $|i \rangle $ is the direct product of the
plane waves with the momenta $ {\bf k}_1 = {\bf p}_1+\frac
{1}{2}{\mbox {\boldmath $\kappa $ }} $ and ${\bf k}_2 = {\bf p}_2
- \frac {1}{2}{\mbox {\boldmath $\kappa $ }}$, while the initial
state $|i' \rangle $ is the direct product of the plane waves
with the momenta ${\bf k}'_1 = {\bf p}_1- \frac {1}{2}{\mbox
{\boldmath $\kappa $ }}$ and ${\bf k}'_2 = {\bf p}_2+\frac
{1}{2}{\mbox {\boldmath $\kappa $ }}$.  Instead of the luminosity
$L$ the number of events starts to depend on the quantity
\begin{equation}
L({\bf r} )= v_{12}\int n_1({\bf r}_2 - {\bf r},t)~n_2({\bf
r}_2,t) \, d^3 r_2 dt
\label{27}
\end{equation}
through the following formula
\begin{equation}
dN = \int \frac {d^3\kappa d^3 r}{(2\pi)^3} \;
\mbox{e}^{ i{\mbox{\scriptsize\boldmath$\kappa$}} {\bf r}} \,
d\sigma ({\mbox {\boldmath$\kappa$}})\, L({\bf r}).
\label {28}
\end{equation}
Note that in this formula the densities $n_1$ and $n_2$ are taken
in different points --- it corresponds to the nonlocal interaction
of colliding muons (contrary, in eq. (\ref{25}) these
densities are taken
 in one and the same point).

The characteristic values of $ \kappa $ are of the order of the
inverse beam sizes, i.e.
$$
\kappa \sim {1\over a}.
$$
Usually this quantity is much smaller than the typical scale for
the variation of the matrix element with respect to the initial
momenta. This is the case for $dN_{{\rm st}}$ which is determined by
the region of large values of $-q^2 > m^2$. Here we can put
$\vkappa =0$ in $d\sigma (\vkappa)$ which immediately results in
the standard expression for the number of events (\ref{25}).

However, for the quantity $dN_{{\rm nst}}$ determined by the region
(\ref{5}), the variable $q^2$ can go through zero. Therefore in
this region $q^2$ changes considerably if initial momenta are
varied on a quantity of the order of $1/a$.  Consequently, 
transformations (\ref {26})--(\ref{28}) reduce to the following
modification:
\begin{equation}
\frac {1}{|q^2 +i\epsilon|^2} \to \frac {1}{t+i\epsilon}~\frac
{1}{t' -i\epsilon}
\label{29}
\end{equation}
where
\begin{equation}
q^2 = (p_1 -p_3)^2,~~~  t = (k_1-p_3)^2 , ~~~ t'= (k'_1-p_3)^2 .
\label{30}
\end{equation}
Let us expand $t$ and $t'$ up to the terms
linear in $\vkappa$. This gives
\begin{equation}
t=q^2- \lambda (q^2) ,~~~ t'=q^2 + \lambda (q^2)
\label{31}
\end{equation}
where
\begin{equation}
\lambda (q^2) = \vkappa {\bf Q}, ~~~ {\bf Q} = -{\bf p}_3 + {E_3 \over
E_1}~ {\bf p}_1.
\label{32}
\end{equation}

As the result, the divergent quantity
$$
B = \int \limits _{-m^2}^{t_0} \; {dq^2 \over |q^2+i \epsilon|^2}
$$
transforms to
\begin{equation}
B = \int ~ \frac {d^3\kappa d^3r}{(2\pi)^3} \;
\mbox{e}^{ i{\mbox{\scriptsize\boldmath$\kappa$}} {\bf r} } \;
\frac {L({\bf r})}{L}~
\int  \limits ^{t_0} _{-m^2} ~\frac {dq^2}
{[q^2- \lambda (q^2)+i\epsilon]~[q^2 + \lambda (q^2) -i\epsilon]}.
\label {33}
\end{equation}
As has been shown in \cite {MS} the main contribution to the
integral over $q^2$ in  eq.~(\ref {33}) is given by the pole
$$
q^2=-\lambda(q^2)+i\epsilon \approx -\lambda (q^2=0) +i\epsilon
$$
in the upper half plane.

In the rest frame of the
$\mu ^-$, the vector ${\bf Q}$ is equal to
$$
{\bf Q} =- {\bf p}_3={\bf q} =\omega {\bf n}
$$
where $\omega$ is the energy of the real ($q^2=0$) neutrino
and ${\bf n}$ is the unit vector which defines
direction of motion of the neutrino.

Taking this pole and performing subsequent integrations
over $\vkappa$ and ${\bf r}$, we obtain
\begin{equation}
B= \frac {\pi}{\omega} \int \limits _{0}^{\infty} \;
{L(r {\bf n})\over L} \; dr.
\label {34}
\end{equation}

As a result,
\begin{equation}
dN_{{\rm nst}}= d\sigma_{\nu \mu \to W} {mx \over \omega}
\, d\Gamma \, \int \limits _{0}^{\infty} \; L(r {\bf n})dr \,.
\label{35}
\end{equation}
For unpolarized muon beams
\begin{equation}
d\Gamma ={\Gamma \over \pi} (1-y)(1+2y)dxdy d\varphi
\label{36}
\end{equation}
where $\varphi$ is the azimuthal angle of the vector $\vq$.

In the rest frame of the $\mu ^-$ we have
\begin{equation}
\omega =\frac {m}{2}(1-y),~~~
x=\frac {qp_2}{p_1p_2}=\frac {1}{2}(1-y)(1+\cos{\theta}).
\label {37}
\end{equation}
Here $\pi - \theta$ is the angle between momentum ${\bf q}$ of the 
muonic neutrino and the momentum ${\bf p}_2$ of the positively 
charged muon.

Using these equations we express differential width of the
$\mu ^-$ as
$$
d\Gamma =
\frac {1}{2\pi\tau} (1-y)^2(1+2y)dyd\cos{\theta} d\varphi.
$$
Now adding all pieces together and using (see also eq. (\ref{21}))
$$
dr d\cos{\theta} d\varphi = \frac {d^3 r}{r^2},
$$
we finally get the following expression
$$
dN_{{\rm nst}} = g(y)dy \int \; v\, {n_1({\bf
r}_2- {\bf r}) \over 4\pi\tau r^2 } \; n_2({\bf r}_2,t)\,
d^3r\, d^3r_2 dt,
$$
\begin{equation}
g(y)=2(1-y)^2(1+2y).
\label{38}
\end{equation}
It is easy to see that after redefinition
$$
y=1- {2\omega \over m}, \;\; {\bf r}_2 - {\bf r} ={\bf r}_1
$$
expression (\ref{38}) completely coincides with  eqs.
(\ref{16})--(\ref{20}).

\section{Remark on the case of a large--size beam}

What happens if conditions (\ref{8})-(\ref{9}) are not
satisfied? Though this question is purely academic one for muon
colliders, it can have a certain principal interest. There is no
general answer in this situation and every such case should be
considered separately. In this situation the method of preparation of
colliding beams as well as decrease in the numbers of muons and
neutrinos during the collision are of great importance. The true
consideration in this case would require the modelling of how
really unstable particle travels in the collider --- it is clear
that some characteristics of the collider are likely to enter the
final result at this point.

In this section we consider a particular case related to
the paper \cite{Ginz}. In this paper the divergent cross section
of the process (\ref {1}) has been regularized by
introducing a complex mass for  initial unstable particles:
$$
m \to m -{i\over 2}\,\Gamma \,.
$$
Solving the energy--momentum conservation conditions it is
possible to show then that the square of the momentum transfer in
the $t$-channel gets small imaginary part:
$$
q^2 \to q^2 -{i\over 2}\, m\, (1-y)\Gamma.
$$
Therefore, the singularity is regularized and the calculation of
the cross section starts to be possible. Below we explain the
physical meaning of the result obtained in this way in
ref. \cite{Ginz} \footnote {See also ref. \cite {Peirles}, where a similar
regularization has been used for the process $\pi N^* \to \pi N^*$.}.

Let us assume that $\mu^-$ beam has been prepared much earlier
than the collision takes place. This means that almost all muons
have managed to decay. In this case every $\mu^-$ produces a
single $\nu_\mu$ with definite distribution over the energy
(\ref{17}) and with the isotropic angular distribution in the
rest frame of the $\mu^-$.  To obtain the results of ref.
\cite {Ginz} we also have to assume that the $\mu^+$ beam is so
large that all neutrinos produced in the $\mu ^-$ decay are still
inside this beam having a possibility to collide with the muons
of the positive charge.  This situation is similar to the fast
decay of the $\mu ^-$ in a medium of the muons of the positive
charge. In such situation the number of events is defined by
the cross section $\sigma_{\mu \mu} $ which corresponds to the
cross section of the process (\ref{2.5}) averaged over the
spectrum of neutrinos and their escape angles:
\begin{equation}
\sigma_{\mu \mu}  = \int \; v \; w(\omega)
d\omega \, {\sin{\theta} d\theta d\varphi \over 4\pi}\; \sigma
_{\nu \mu \to W}(s_{\nu \mu})
\label{41}
\end{equation}
where $v$ is defined in (\ref{20}). Using eq. (\ref{37}) and
integrating over $y$ and $\varphi$  we obtain
\begin{equation}
\sigma_{\mu \mu} = \int \limits _{0}^{1}  \; 4x(1-x) (2-x)
\; \sigma _{\nu \mu \to W}(xs)\, dx.
\label{42}
\end{equation}
This result is identical to the one which has been presented in
ref. \cite {Ginz}.

\section {Conclusions}

1. Comparing the results of  sections 3 and 4, we conclude that {\it
the linear beam--size effect obtained in} \cite{MS} {\it
corresponds to the scattering of the $\mu^+$ beam on the
``cloud'' of real neutrinos $\nu_\mu$ produced in the $\mu^- \to
e \bar \nu_e \nu_\mu$ decay}. Such interpretation  assumes that
the colliding beams have finite sizes and that decays of the
negative muons on their way to the interaction point produce a
$\nu_\mu$ cloud which fills the region occupied by $\mu^+$ beam.

Really amazing is the fact that within perturbative calculations
these effects can be found only if one considers a collision of
two wave packets which correspond to the beams of the colliding
particles. Standard approximations in the scattering theory
do not work in this case.

It is interesting to estimate the ``formation length'' $l_f$ of
this $\nu _\mu $ cloud. In the center of mass frame
the angular spread of the muonic
neutrinos $\vartheta$ is of the order of $\vartheta \sim 1/\gamma $.
Hence one obtains
\begin{equation}
l_f \sim {a\over \vartheta} \sim \gamma  \, a.
\label{39}
\end{equation}
For the discussed parameters of the beams \cite {Palm}
and the total energy $100$ GeV one finds $l_f \sim
1$ cm.

From this we  obtain the number of the neutrinos
$\nu_\mu$ which participate in the collision. It can
be estimated as
\begin{equation}
N_\mu \; {l_f \over c \gamma_\mu \tau}  \sim
N_\mu \; {a \over c \tau} \sim 10^{4}
\label{40}
\end{equation}
for $N_\mu \;  \sim 10^{12}$. This number shows the level of
statistical fluctuations which one expects for the $\nu_\mu
\mu^+$ collisions.

2. It is interesting to compare the process (\ref{1}) which has been
discussed in this paper
and the processes with large impact parameters considered in
the review \cite{Obzor}, for example, the process
\begin{equation}
ep \to ep\gamma.
\label{43}
\end{equation}
It is not difficult to show that the impact parameters
$\varrho$, which give essential contribution to the cross
section of the process (\ref {43}) in the standard calculations,
are of the order of
$$
\varrho \stackrel {<}{\sim} \; \varrho_m \sim {1\over m_e}
{4E_eE_p(E_e-E_\gamma) \over m_e m_p E_\gamma}.
$$
Note that the quantity $\varrho _m$ can be macroscopically large:
for instance,
for the collider HERA $\varrho_m \stackrel {>}{\sim} 1$ cm already
for $E_\gamma \stackrel {<}{\sim} 0.3$ GeV. In this case there is
a cloud of equivalent photons (EP) around every proton, these
EP scatter on the electrons of the opposite beam.
As a result, the
interaction of electrons with  protons becomes {\it macroscopically }
nonlocal. In the
discussed process (\ref{1}) similar macroscopically nonlocal
interaction between  $\mu^+$ and $\mu^-$ is due to the cloud of
real $\nu_\mu$'s which appears as a result of the muon decay.
Certainly, the nature of these clouds is quite different,
however, one can describe both processes by means of the
same technique.

Distributions of EP and $\nu_\mu$'s in $\varrho$ are
absolutely different for both of these processes.
The $\nu_\mu$ cloud in the process (\ref{1}) is distributed
in the impact parameter space according to
eq. (\ref {13}), i.e.
$$
dN_\nu \propto {d^2\varrho\over \varrho}.
$$
This distribution leads (after integration over
$\varrho$ in the beam region, $\varrho \stackrel {<}{\sim} a$) to
the linear beam--size effect:
$$
d\sigma_{\mu \mu} \propto a.
$$
The distribution of EP has another form
\begin{equation}
dN_{{\rm EP}} \propto {d^2 \varrho \over \varrho^2}
\label{44}
\end{equation}
from which the known logarithmic beam--size effect immediately follows:
\begin{equation}
d\sigma_{ep} \propto \ln{a}.
\label{45}
\end{equation}
Let us emphasize that the logarithmic beam--size effect has been
already established in
experiments \cite{IYAF,HERA}.

3. Let us finally compare the contribution of
real and virtual $\nu_\mu$'s to the total cross section.
First, we remind that because of the fixing of the region of the
final phase space, the cross section
$\sigma_{{\rm st}}$ (see eq. (\ref{4}) and the discussion after it)
is completely determined by virtual
$\nu_\mu$'s with $q^2 < -m^2$. Contrary, the contribution of real
$\nu_\mu$'s in the intermediate state absolutely dominates in
$\sigma_{{\rm nst}}$ -- the relative contribution of virtual
neutrinos to this piece of the cross section can be estimated as
$\sim 1/(a\sqrt{t_0}) \sim 1/(am) \sim 10^{-10}$.  Let us stress,
however, that the approach described in \cite {MS} allows to
calculate the contribution of virtual neutrinos to $\sigma_{{\rm
nst}}$ as well as the contribution which appears due to the
interference of real and virtual neutrinos.

\section*{Acknowledgments}

We are indebted to I.~F.~Ginzburg, V.~S.~Fadin, D.~Yu.~Ivanov,
V.~V.~Serebryakov and  V.~M.~Strakhovenko for useful discussions.
K.M.  is grateful to the Graduiertenkolleg
``Elementarteilchenphysik'' of the Mainz University for the
support. G.L.K and V.G.S.  acknowledge the support of the Russian
Fund of Fundamental Research.


\begin{thebibliography}{99}
\bibitem{Ginz}
I.F.~Ginzburg, Preprint DESY {\bf 95-168} (1995) and
hep-ph {\bf 9509314}, (unpublished).
\bibitem{MS}
K.~Melnikov and V.G.~Serbo, Phys. Rev. Lett. (in print) and 
hep-ph {\bf 9601221}; Preprint MZ-TH-96/03 and
hep-ph {\bf 9601290}.
\bibitem{Klass}
G.L.~Kotkin, S.I.~Polityko and V.G.~Serbo, Yad. Fiz.  {\bf 42},
692~(1985).
\bibitem{Obzor}
G.L.~Kotkin, V.G.~Serbo and A.Schiller, Int. Journ.
Modern Physics {\bf A7}, 4707~(1992).
\bibitem{Palm}
D.~Cline, Nucl. Instrum. Methods {\bf 350}, 24 (1995);
R.B.~Palmer,  Beam Dynamics Newsletters, {\bf 8} (1995).
\bibitem{Okun} L.~B.~Okun, {\it Leptons and Quarks}
(North-Holland, Amsterdam, 1982).
\bibitem{LL}
L.D.~Landau and E.M.~Lifshitz, Classical Theory of Fields,
Pergamon, Oxford, 1962.
\bibitem{IYAF}
A.I.~Blinov et al, Phys.Lett. {\bf B113}, 423~(1982);
V.N.~Baier, V.M.~Katkov and V.M.~Strakhovenko, Yad. Fiz. {\bf
36}, 163 (1982);
A.I.~Burov and Ya.S.~Derbenev, Preprint 81-64 INP (Novosibirsk,
1981), (unpublished).
\bibitem{HERA}
K.~Piotrzkowski, Zeit. f. Phys. {\bf C67}, 577~(1995).
\bibitem{Peirles} 
R.~F.~Peierls, Phys. Rev. Lett. {\bf 6}, 641 (1961).
\end{thebibliography}
\end{document}